\documentclass[12pt]{article}
\usepackage{epsfig,amsfonts,amssymb}
\usepackage{hyperref}
\pdfoutput=1

\usepackage{cite}
\usepackage{cite}

\usepackage{comment}

\def\ZZZ{{\hbox{ Z\kern-1.6mm Z}}}
\def\RRR{{\hbox{ R\kern-2.4mm R}}}
\def\CCC{{\hbox{ C\kern-2.0mm C}}}
\def\zzz{{\hbox{z\kern-1mm z}}}

\def\ZZZ{\mathbb{Z}}
\def\RRR{\mathbb{R}}

\newcommand{\qeq}{{\hbox{=\kern-2.3mm ? \kern.5mm }}}
\renewcommand{\qeq}{=}

\newcommand{\eps}{\epsilon}

\newcommand{\MM}{{\cal M}}

\newcommand{\wt}{\widetilde}

\newcommand{\be}{\begin{equation}}
\newcommand{\ee}{\end{equation}}
\newcommand{\ben}{\begin{eqnarray}\displaystyle}
\newcommand{\een}{\end{eqnarray}}

\newcommand{\refb}[1]{(\ref{#1})}

\newcommand{\sectiono}[1]{\section{#1}\setcounter{equation}{0}}

\def\one{{\hbox{ 1\kern-.8mm l}}}
\def\zero{{\hbox{ 0\kern-1.5mm 0}}}

\newcommand{\bea}[1]{\begin{eqnarray}\label{#1} }
\newcommand{\eea}{\end{eqnarray}}

\newcommand{\eqref}{\refb}

\usepackage{bm}
\usepackage[table]{xcolor}

\begin{document}

\baselineskip 24pt

\begin{center}

{\Large \bf How to Expose a Black Hole}

\end{center}

\vskip .6cm
\medskip

\vspace*{4.0ex}

\baselineskip=18pt

\centerline{\large \rm Ashoke Sen}

\vspace*{4.0ex}

\centerline{\large \it International Centre for Theoretical Sciences - TIFR 
}
\centerline{\large \it  Bengaluru - 560089, India}


\vspace*{1.0ex}
\centerline{\small E-mail:  ashoke.sen@icts.res.in}

\vspace*{5.0ex}

\centerline{\bf Abstract} \bigskip

According to the correspondence principle of Horowitz and Polchinski, many
black holes in string
theory are continuously deformed to usual quantum systems involving D-branes and fundamental
strings when the string coupling becomes sufficiently small. Therefore if we consider a configuration in
space-time where the dilaton varies over an appropriate range, then a black hole moving in such a
background will smoothly transition from the black hole state to a normal quantum state whose
microstates are not hidden behind an event horizon. The possible obstruction to this mechanism comes from
the fact that if the dilaton varies too fast then 
the adiabatic approximation may break down and / or 
the ambient space-time itself may collapse to a black hole
and get hidden from the asymptotic observer. On the other hand,
if the dilaton varies too slowly then the time that
it takes for the black hole to travel the required distance will exceed the evaporation 
time of the black hole.
We show that by choosing the background appropriately these obstructions can be avoided and a
gentle motion towards the weak coupling region will convert the black hole into a normal quantum
state without an event horizon.

\vfill \eject

\tableofcontents

\sectiono{Introduction} \label{s1}

Our conventional picture of black hole formation and subsequent evaporation takes an
asymmetric view: the black holes form by collapse of ordinary matter by classical process, but
their evaporation happens by quantum process in the form of Hawking radiation, usually taking much
longer than the time it takes for the formation. The goal of this paper will be to argue that in
certain class of string theories we can partially avoid this asymmetry: A classical process can
convert the black hole to a regular state without event horizon 
that can be probed by an external observer.
Furthermore, the time needed to convert the black hole to a regular quantum state can be made
parametrically small compared to what would be the evaporation time of the black hole if left
unattended. 

The main idea will be to use the 
correspondence principle formulated in \cite{9612146,9707170,9907030}, 
following earlier suggestions in
\cite{bowick,9309145,9605112,9609075}. More recent perspective on this can be found in
\cite{2109.08563,2107.09001,2205.15976,2307.03573}. This principle
states that as we adiabatically change the string coupling
from finite value to a sufficiently 
small value, a Schwarzschild black hole transforms to a highly excited
elementary string state. This was generalized to a wide class of other charge
carrying black holes, although at weak coupling these black holes correspond to
systems of D-branes and fundamental strings instead of just fundamental strings.

Recently it was shown in \cite{2502.07883,2503.00601,2506.13876} 
(see \cite{2501.17697} for a different viewpoint)
that in string theories with moduli fields in asymptotically flat space-time, 
one can produce
finite energy background in which we have an arbitrarily large region in space-time inside which
the moduli can take any desired values, in general different from their asymptotic values.
More generally one can produce backgrounds in which the moduli vary from any desired value
to any other, and the variation can be made arbitrarily slow.

This raises the possibility of realizing the black hole to elementary string / D-brane 
transition dynamically. Given a black hole,
if the transition to the microscopic description happens around a value $g_0$ of the string coupling
(which is a modulus of the theory) then by producing a background in which the coupling varies from
a value much larger than $g_0$ to a value much smaller than $g_0$, and making the
black hole roll in this background, we can achieve the desired transition. 
Once the black hole rolls to the weak coupling region, it is
described by an ordinary quantum system. While determining the exact quantum state of the system
may still be a challenging
problem, it is no different from a high energy pure quantum state of matter
that looks approximately thermal.

There are however some consistency conditions that need to be checked. While we 
can produce backgrounds in which the moduli 
vary arbitrarily slowly in space-time, there is an upper bound on how
fast the moduli can vary. This comes from the fact that too fast a variation will make the
background space-time collapse into a big black hole, and the whole system will get
hidden behind an event horizon. Hence we have a lower limit on the time it takes
for the black hole to travel from a region where the black hole description is valid to the region
where the microscopic description is valid. Other lower bound on the 
transit time comes from the fact that in order
that the adiabatic description is valid, the background must vary slowly
in the frame of the black hole. At the same time, the time
of travel must be small compared to the evaporation time of the black hole. We show that it is
possible to satisfy all of these conditions together. Hence in such a background, a gentle roll
towards the weak coupling region will convert the black hole to a regular quantum system whose microstates are not hidden behind an event horizon.

A similar mechanism for converting a black hole to an elementary string was considered in \cite{2307.03573}
with the help of large amplitude and large wavelength dilaton wave. However, as discussed in \cite{2506.13876},
there is an upper bound of order unity on the amplitude of a scalar wave in any dimensions, since for 
a wave of amplitude $a$ and wave-length $R$ measured in Planck units, 
the energy density is of order $(a/R)^2$ 
and the energy contained in a ball of size $R$ in $D$
space-time dimensions is of order $a^2R^{D-3}$. The associated Schwarzschild radius $a^{2/(D-3)}R$
will exceed $R$ for sufficiently large $a$, making the background collapse into a 
black hole.\footnote{This is not
a no go theorem since this argument does not take into account the full non-linear effects of gravity. However this
shows that without having a more detailed analysis, we cannot trust the dilaton wave background to be
horizon free.}
The background 
considered
here does not suffer from this problem. Even though we use a large black hole to produce the desired background,
the system that we study always remains outside the horizon of the large black hole and hence remains accessible
to the asymptotic observer.

Note that the analysis of this paper is agnostic about the various solutions to the black information
paradox\cite{Hawking:1976ra,Page:1993wv} 
that have been discussed in the literature. For example in the fuzzball 
proposal\cite{0502050,0909.1038} even at
finite string coupling the horizon is replaced by horizonless geometry. The mechanism discussed in this
paper would convert these horizonless geometries to regular weakly coupled quantum system made of
strings and branes. In the island 
proposal\cite{Penington:2019npb,Almheiri:2019psf,1908.10996,Almheiri:2020cfm} 
for an old black hole the degrees of freedom inside the
black hole are encoded in the outgoing Hawking radiation. Once the black hole
rolls  across the correspondence point, the horizon shrinks away and with that the island
also disappears. In the holography of information proposal 
\cite{Laddha:2020kvp} one determines the quantum state of the
black hole from the asymptotic region with the help of gravitational constraint equations. Just as we
do not need to invoke this for finding the quantum state of an ordinary hot coal, similarly once the
black hole crosses the correspondence point, one can study this just as we shall study an ordinary
hot coal.

The rest of the paper is organized as follows. In section \ref{s2} we show how we can produce a
background in which a Schwarzschild black hole can roll from the black hole state to a 
highly excited elementary string state. In section \ref{s3} we briefly discuss the case of other
non-extremal black holes. In section \ref{s4} we discuss the case of BPS black holes. We end in
section \ref{s6} with some comments.

\sectiono{Schwarzschild black hole} \label{s2}

In this section we shall describe the dynamical transition of a Schwarzschild black hole to an
elementary string excitation. In section \ref{s2.1} we review the correspondence principle
for this system. In section \ref{s2.2} we describe the general strategy for producing the
background that can be used to dynamically change a black hole to an elementary
string state. In section \ref{s2.3} we describe explicit realization of such a background.
In section \ref{s2.4} we describe an extension of the analysis in section \ref{s2.3} where
the string coupling varies from a value of order unity all the way to a sufficiently small
value where the black hole becomes an elementary string state.

\subsection{Review of the correspondence principle} \label{s2.1}

Let us work in a string compactification where we have $D$ non-compact space-time
directions and let us denote by $g_s$ the string coupling in $D$ dimensions. We shall
keep the size of the compact directions finite in string units and hence the string coupling
in ten dimensions is also of order $g_s$.
Let $\wt m$ be the mass of any state measured in string units and $m$ be the mass of the
same state
measured in $D$ dimensional Planck units. Newton's gravitational
constant is, by definition, of order unity in Planck units and is of order $g_s^2$ in string
units. Then 
we have the relation:
\be \label{etr1}
m \sim g_s^{2/(D-2)}\, \wt m\, .
\ee
Similarly, if $\wt r$ denote the size of the system measured in string units and $r$ denotes the same
size measured in Planck units, then we have the relation
\be \label{etr2}
r \sim g_s^{-2/(D-2)} \, \wt r\, .
\ee
In most of this section we shall use Planck units to express various quantities, but we can translate
the results to string units using \refb{etr1} and \refb{etr2}.

Let us consider an elementary string state at level $N$. For large $N$,
the entropy of these states, defined as the logarithm of the degeneracy, varies as
\be \label{e1}
S \sim \sqrt N\, .
\ee
The mass of the state in string units is given by
\be\label{e2a}
\wt m  \sim \sqrt N \, ,
\ee
which translates to
\be \label{e2}
m \sim g_s^{2/(D-2)} \,   \sqrt N
\ee
in Planck units.

Let us now leave this system aside and consider a Schwarzschild or Kerr
black hole of mass $m$ in Planck units. 
The radius $r_s$ of the black hole measured in Planck units is given by
\be \label{ers}
r_s \sim m^{1/(D-3)} \, .
\ee
The entropy of such a black hole is of order
\be\label{e3}
S \sim r_s^{D-2}\sim m^{(D-2)/(D-3)} \, .
\ee
The 
temperatures $T_{bh}$ of the black hole, measured  in  Planck units, is given by:
\be \label{etbh}
T_{bh}\sim m^{-1/(D-3)}\, .
\ee
The
evaporation time $\tau_{bh}$
of the black hole, measured in the Planck
units, is obtained by dividing the total mass by the
product of the area of the horizon $\sim r_s^{D-2}$ and the energy flux $\sim T_{bh}^D$.
This gives:
\be \label{etimebh}
\tau_{bh} \sim m^{(D-1)/(D-3)}\, .
\ee

According to the correspondence principle\cite{9612146,9707170},
at some particular value of $g_s$, the description of the system changes from that of a
black hole to that of a fundamental string. 
The value of $g_s$ where it happens is determined by demanding that at 
the transition point  the entropy and the mass of the black hole should match
those of the elementary string states. Comparing \refb{e1},
\refb{e2} with \refb{e3}, we see that this gives
\be
\sqrt N \sim g_s^{2/(D-3)} N^{(D-2)/\{2(D-3)\}}\, , 
\ee
and hence
\be\label{e2.9}
g_s \sim N^{-1/4}\, .
\ee
Using  \refb{e2}, \refb{ers} and \refb{e2.9} we see that at this point
\be \label{e2.9a}
 m \sim N^{(D-3) / \{2 (D-2)\}}, \qquad  r_s\sim N^{1/\{2 (D-2)\}}\, .
\ee
Using \refb{etr2}, \refb{e2.9} and \refb{e2.9a} 
we now see that the radius of the black hole, measured in
string units, is given by,
\be\label{erstring}
\wt r_s \sim 1\, .
\ee
Also from \refb{etbh}, \refb{etimebh}, \refb{e2.9} and \refb{e2.9a} 
we see that at the transition point
\be \label{e2.14}
T_{bh} \sim N^{-1/\{2(D-2)\}},  \qquad
\tau_{bh}  \sim N^{(D-1)/\{2(D-2)\}}\, .
\ee
Using \refb{etr1} and \refb{e2.9}, and using the fact that the conversion formula for the
temperature is the same as that of the mass, we get the temperature of the black hole
in string scale
\be \label{etempstring}
\wt T_{bh}\sim 1\, .
\ee
From \refb{erstring} and \refb{etempstring} we see that 
at the transition point the size of the horizon and the temperature of the
black hole reaches the string scale, confirming that stringy effects become important
at this point.

\subsection{Dynamical transition from black hole to string} \label{s2.2}

Let $\phi$ denote the dilaton field. We restrict our analysis to those string compactifications where
$\phi$ is a modulus.
Our goal will be to create a  background in which $e^\phi$, that
determines the local string coupling
$g_s$, changes from $N^{-1/4}/\eps$ to $N^{-1/4}\eps$ for some small but fixed number $\eps$.
The distance scale over which this happens should be much larger than the\
size of the black hole / string system we have analyzed in section \ref{s2.1},
so that the black hole can be treated as a point particle moving in this
background. Furthermore the various fields must vary sufficiently slowly so that 
the black hole moving through this background locally sees a configuration that is close
to the vacuum. In particular the effect of any
ambient radiation / other sources of stress tensor on the black hole should
remain small, and the
motion of the black hole can be taken to be adiabatic.
Then as the black hole traverses this region, it will change into a fundamental string state whose
internal state is not hidden from the asymptotic observer behind an event horizon.
At the same time the size of the region over which the transition takes place cannot be
arbitrarily large since the time taken for this process must be 
parametrically smaller than the evaporation time so that the black hole does not
change appreciably during the transit.
Furthermore, 
the energy of the required background should not be so high that it gets
hidden behind an event horizon.
Our goal will be to show that we can produce such a background. 

We shall first describe the general strategy for achieving this and then illustrate this using
a specific background. 
During this analysis we shall work with quantities measured in Planck units, but the final
results of course will be independent of the units used in the analysis. 
Let us suppose that we have a (possibly time dependent)
solution to the classical equations of motion in which we have a time-like trajectory along
which $e^{\phi}$ 
changes from $N^{-1/4}/\eps$ to $N^{-1/4}\eps$, and the proper time along the trajectory
does
not scale with $N$ (when expressed in Planck units). This is a reasonable assumption since the
net change in the canonically normalized field $\phi$ is independent of $N$. 
In section \ref{s2.3} we shall construct an explicit background satisfying this requirement and in 
section \ref{s2.4}
we shall generalize this analysis to the case where $e^{\phi}$ 
changes from a finite value to $N^{-1/4}\eps$.
We shall further assume
that the beginning and the end points of the trajectory can send signals to the asymptotic observer.
Then we can generate another solution to the classical equations
of motion by scaling all covariant rank $k$
tensor fields by $\lambda^k$ and contravariant rank $k$ tensor fields by $\lambda^{-k}$, since under
such a transformation a two derivative action scales by $\lambda^{D-2}$ in $D$ space-time
dimensions and the equations of motion remain unchanged\cite{9707207,2502.07883,2503.00601,2506.13876}. In particular the $\sqrt{\det g}$ factor
in the Lagrangian density produces a factor of $\lambda^D$ and the extra factor of $g^{\mu\nu}$ that
contracts with the two derivatives produce a factor of $\lambda^{-2}$.
Furthermore, the causal structure of space-time remains unchanged under this rescaling and the
beginning and the end points of the trajectory can still send signals to the asymptotic observer.
Since the metric scales by $\lambda^2$,  
the proper time taken during the travel scales as $\lambda$ and 
becomes large in the limit of large $\lambda$. 
In order to keep this small compared to the evaporation time $\tau_{bh}$
given in \refb{e2.14}, we need
\be\label{efirst}
\lambda <<  N^{(D-1)/\{2(D-2)\}}\, .
\ee
Note that on the right hand side of \refb{efirst}
we have used the evaporation time at the crossover point and not taken into
account the effect of the small parameter $\eps$. This is because in the $<<$ symbol it will 
be assumed implicitly that the inequality will hold even when the right hand side is multiplied
by some appropriate power of $\epsilon$ to represent the worst case scenario.

\refb{efirst} gives an upper bound on $\lambda$. 
We shall now derive a lower bound by requiring
that the change in the state of the  black hole is adiabatic during the transition.
If the black hole moves at a speed of the order of the speed of light, then the distance covered
per unit time is of the order of the radius of the black hole. We shall require that the 
background does not change appreciably over this distance. 
Since the distance over which the background changes scales as $\lambda$, this imposes
the condition, using \refb{e2.9a},
\be\label{emain}
\lambda >> r_s \sim N^{1/\{2 (D-2)\}}\, .
\ee
Under this condition the rate of change in
the background fields, as seen from the perspective of the black hole, remains small and the
transition can be called adiabatic.
As a consequence of \refb{emain} the system also satisfies some other conditions that
we shall now describe.
\begin{enumerate}
\item Since the length scale over which the background changes  
is  large compared to the size of the black hole
/ string, we can describe the motion  as that of a point particle moving in a background. 
\item The acceleration
of the black hole along the trajectory scales as $\lambda^{-1}$ and hence
the Unruh temperature\cite{Unruh:1976db} seen by the black hole scales as 
$\lambda^{-1}$.\footnote{Note that this is a crude estimate
based on scaling and the actual acceleration and the 
associated temperature will depend on the details of the trajectory.
While free fall along a geodesic provides a natural trajectory, this is not needed.
For example the black hole may be gravitationally bound to another set of objects fitted
with engines 
that slows down the fall.  \label{fo1}}
\refb{emain} ensures that this is small compared to the black hole temperature
$ N^{-1/\{2(D-2)\}}$ given in \refb{e2.14}.
\item 
Since the local Unruh temperature is small compared to the black hole temperature,
the total energy absorbed by the black hole during the transit is small compared to the
total energy emitted by the black hole during the transit.
The latter, in turn, is already small compared to the total mass
of the black hole due to \refb{efirst}. Therefore the energy absorbed by the black hole during
the transit remains small compared to its total mass.
\item A similar argument shows that 
the entropy absorbed by the black hole from the ambient radiation during
the transit also remains small compared to its total entropy.
\end{enumerate}
From this analysis we conclude that as long as \refb{efirst} and 
\refb{emain} hold, the change in the state
of the black hole can be taken to be adiabatic. Combining \refb{efirst} and \refb{emain} we
get,
\be\label{eboth}
N^{1/\{2 (D-2)\}}  << \lambda << N^{(D-1)/\{2(D-2)\}} \, .
\ee

This shows that as long as we can create a background in which the string coupling
varies from $N^{-1/4}/\eps$ to $N^{-1/4}\eps$ for some small number $\eps$ that does not scale 
with $N$, we can scale the solution and by choosing the
scaling parameter appropriately, we can ensure that the transition from black hole to elementary
string is adiabatic and the total
fractional change in the energy and entropy of the black hole during the transit remains small.
The ability to produce such a varying dilaton configuration was discussed in
\cite{2502.07883,2503.00601,2506.13876}
where it was shown that starting from any set of asymptotic 
values of the moduli, we can produce any other set of values inside an arbitrarily
large region of space-time.
Hence we can utilize the construction of \cite{2502.07883,2503.00601,2506.13876} 
and then scale it by a parameter $\lambda$
satisfying \refb{eboth} to produce a background  that will
change a black hole to an elementary string excitation. 
Once the system becomes an excitation of an elementary string, it
can be probed by an external observer like any other quantum state.

Let us now discuss in what sense the state of the black hole does not  change during the
transit. Since the black hole radiates and absorbs radiation during the transit, the quantum
state of the black hole itself changes during this process. To prevent this we would have to
demand that the black hole does not emit or absorb even a single quantum of
radiation during the transit, which
will be too strong a condition to hold. For example if we demand that not even a single Hawking
quanta is emitted by the black hole during the transit, or equivalently that the total entropy of 
the radiation emitted during the transit is small compared to unity, we get the condition
\be
\lambda\times T_{bh}^{D-1} \times r_s^{D-2} << 1 \qquad \Rightarrow \qquad 
\lambda <<  N^{1 / \{2 (D-2)\}}\, .
\ee
This is in conflict with the lower bound on $\lambda$ given in \refb{eboth}. Indeed,
since $r_s\sim N^{1/\{2(D-2)\}}$, 
this would imply that the transit time $\lambda$ is small compared to the size of the
black hole which is clearly not compatible with the adiabaticity of the transition.
However, some properties of the state  remain unchanged during the transit.  
For example, if at the beginning of the roll the back hole was formed out of the collapse of a pure state, then
at the end of the roll it will remain almost pure, in that its entanglement entropy will remain small compared
to its thermodynamic entropy since due to the condition
\refb{efirst}, the entropy emitted or absorbed by the black hole during the transit is 
parametrically smaller than $N^{1/2}$.
On the other hand, if the black hole under consideration is an old black hole that is
entangled with the Hawking radiation emitted in the past with an entanglement entropy of
order $N^{1/2}$, then  the entanglement entropy of the black hole
does not change appreciably during the transit and at the end of the transit
we shall get a regular  quantum state of an elementary string
without event horizon,  entangled with the past Hawking radiation with an entropy of order $N^{1/2}$.


\subsection{Varying dilaton background from charged black hole} \label{s2.3}



We shall now describe an explicit example of a background of the type described above.
This will be done with the help of a bigger black hole that we shall call the background
black hole.
For this we shall use the electrically charged black hole solution in heterotic or type IIA
string theory reviewed in
appendix \ref{sa}. The relevant part of the solution takes the form:
\ben \label{ecanmet}
ds^2 &=& - (1+ C\rho^{3-D})^{-2(D-3)/(D-2)} (1-2m\rho^{3-D}) dt^2 + 
(1+ C\rho^{3-D})^{2/(D-2)} (1-2m\rho^{3-D})^{-1}
 d\rho^2 \nonumber \\
&+& (1+ C\rho^{3-D})^{2/(D-2)}\rho^2 d\Omega_{D-2}^2\, , \nonumber \\
e^{-2\phi} &=&  (1+ C\rho^{3-D}) 
\, , \nonumber \\
A_t &=& -{1\over \sqrt 2} \, m\, \sinh\beta\, \rho^{3-D}\, (1+ C\rho^{3-D})^{-1}\, .
\een
Here $\phi$ is the dilaton field, $A_\mu$ is a gauge field that couples to winding + momentum
charge along a compact circle and $ds^2$ is the canonical Einstein frame metric. $m$ and $\beta$ are
independent parameters, and
\be
C = m (\cosh\beta-1)\, .
\ee
We shall take $\beta$ to be large so that $C>>m$ and work in the region
\be\label{e2.31}
m << \rho^{D-3} << C\, .
\ee
In this region we have $m\sinh\beta \simeq C$ and the background black hole solution
takes the form
\ben \label{ebackgroundlimit}
ds^2 &\simeq & - ( C\rho^{3-D})^{-2(D-3)/(D-2)} dt^2 + 
(C\rho^{3-D})^{2/(D-2)}
 d\rho^2 + (C\rho^{3-D})^{2/(D-2)}\rho^2 d\Omega_{D-2}^2 \, , \nonumber \\
e^{-2\phi} &\simeq &  (C\rho^{3-D}) 
\, , \nonumber \\
A_t &\simeq & -{1\over \sqrt 2} \,  (1 - C^{-1} \rho^{D-3})\, .
\een
Note that in the expression for $A_t$ we have kept the leading non-constant term since
the constant term gives vanishing field strength. Since the horizon of the black hole is at
$\rho^{D-3}=2m$, all points in the range \refb{e2.31} are outside the horizon and remain
causally connected to the asymptotic observer.

Now let us consider the effect of scaling by $\lambda$ described in section \ref{s2.2}. Since
the metric scales by $\lambda^2$, the gauge field scales by $\lambda$ and the dilaton remains
constant, we see that we can achieve this by simply scaling $\rho$ and $t$ by $\lambda$ and
$C$ by $\lambda^{D-3}$. Since the scaling of $\rho$ and $t$ are coordinate transformations, this
means that we need to take $C$ to be large. 

We can simplify the solution by introducing new coordinates $\bar\rho$ and $\bar t$ via
\be \label{e2.33}
\bar \rho = C\, \rho, \qquad \bar t = C^{3-D}\, t\, .
\ee
In these coordinates the solution takes the form\footnote{Since $F_{\bar \rho \bar t}$ has an extra
factor of $C^{-1}$ and $C$ is large, one might wonder why we keep this term. The reason is that
when we use the canonical metric, the gauge kinetic term is multiplied by a factor of 
$e^{-4\phi/(D-2)}$. This produces a factor of $C^2$ due to $C$ dependence of $\phi$ and
cancels the $C^{-2}$ factor coming from the quadratic term in the gauge field.}
\ben \label{ebackgroundNEW}
ds^2 &\simeq & -  \bar \rho^{\, 2(D-3)^2/(D-2)} d\bar t^2 + 
\bar\rho^{\, -2(D-3)/(D-2)}
 d\bar\rho^{\, 2} + \bar \rho^{\, 2/(D-2)} d\Omega_{D-2}^2 \, , \nonumber \\
e^{-2\phi} &\simeq &  C^{D-2}\, \bar \rho^{\, 3-D}
\, , \nonumber \\
F_{\bar \rho \bar t} &\simeq & {1\over \sqrt 2} \, C^{-1}\,  (D-3) \, \bar \rho^{\, D-4}\, .
\een
In this form the metric is $C$ independent and the scaling of the metric by
$\lambda^2$ is generated by the
transformation
\be\label{erhobarscale}
\bar\rho \to \lambda^{D-2}\bar\rho, \qquad \bar t\to \lambda^{1 - (D-3)^2}\bar t\, .
\ee
In order to get $e^{\phi}\sim N^{-1/4}$,
we need to take $\bar\rho\sim \bar\rho_0$ where
\be\label{ecdrel}
C^{D-2}\, \bar\rho_0^{3-D}\sim N^{1/2}\, .
\ee
In particular to have the coupling change from $N^{-1/4}/\eps$ to $N^{-1/4}\eps$ for some
small number $\eps$, we need to
have $\bar\rho$ change from $\bar\rho_0\eps^{-2/(D-3)}$ to $\bar\rho_0 \eps^{2/(D-3)}$.
In order to verify that $\rho\sim \bar\rho_0/C$ lies in the range \refb{e2.31}, we note that
according to \refb{ecdrel}, $\rho^{D-3}=(\bar\rho_0/C)^{D-3}\sim C \, N^{-1/2}$ and hence for large $N$
the upper bound $\rho^{D-3}<<C$ is satisfied automatically. On the other hand the 
lower bound $\rho^{D-3}>>m$ may be satisfied by requiring
\be \label{embound}
m<< C\, N^{-1/2}\, .
\ee

Consider now the smaller black hole / string considered in section \ref{s2.1} and \ref{s2.2}
rolling in this background black hole from $\bar\rho\sim\bar\rho_0\eps^{-2/(D-3)}$ 
to $\bar\rho\sim \bar\rho_0 \eps^{2/(D-3)}$. In order to distinguish it from the background black hole,
we shall call this the rolling black hole.
Since the background black hole is much larger than the
rolling black hole, we can treat the latter as a point mass moving in the field of the
background black hole. During the transit the string coupling seen by the rolling black hole changes
from $N^{-1/4}/\eps$ to $N^{-1/4}\eps$, as required for the black hole to string transition.
It follows from \refb{ebackgroundNEW} that the time taken for the roll, as measured
by a local observer who is
stationary with respect to the background,
 scales
as $\bar\rho_0^{1/(D-2)}$
where we have ignored factors containing powers of $\eps$. Since the rolling black hole
speed is not ultra-relativistic, the proper time measured by the rolling black hole also
scales as $\bar\rho_0^{1/(D-2)}$.
Requiring this to be smaller than the evaporation time $\tau_{bh}\sim N^{(D-1)/\{2(D-2)\}}$ 
given in \refb{e2.14}, we get
\be
\bar\rho_0<< N^{(D-1)/2}\, .
\ee
On the other hand, for the adiabatic approximation to hold, the time of transit $\bar\rho_0^{1/(D-2)}$
should be large compared to
the size $r_s\sim N^{1/\{2 (D-2)\}}$ of the horizon given in \refb{e2.9a}. This gives
\be \label{e2.36A}
\bar\rho_0^{1/(D-2)}  >> N^{1/\{2(D-2)\}} \qquad \Rightarrow \qquad  \bar\rho_0 >> N^{1/2}\, .
\ee
Combining the two bounds, we get
\be \label{erhorange}
N^{1/2} << \bar\rho_0 << N^{(D-1)/2}\, .
\ee
\refb{ecdrel} now gives
\be \label{e2.40}
N^{1/2} << C << N^{(D-2)/2}\, .
\ee
This shows that in order to see the black hole to string transition for the rolling black hole,
the background black hole
parameter $C$ needs to be constrained.

Using these results 
we can also test the other conditions. For example using the lower bound on $C$ given
in \refb{e2.40} 
we can verify that the mass of the background black hole, which is of order
$C$, is much larger than the mass of the rolling black hole which is of order $N^{(D-3)/\{2(D-2)\}}$.
Hence the backreaction of the rolling black hole on the background black hole geometry
remains small. In particular, the causal structure of space-time remains unaffected and as
long as the rolling black hole stays outside the horizon of the background black hole, given
by the condition $\rho^{D-3}>2m$, it will be causally connected to the asymptotic observer.
Also,
it follows from
\refb{ebackgroundNEW}
that for $\bar\rho\sim\bar\rho_0$, any
scalar constructed from the fields and two derivatives scales as $\bar\rho_0^{-2/(D-2)}$.
The only exception is $F_{\mu\nu} F^{\mu\nu}$, but the combination $e^{-4\phi/(D-2)}
F_{\mu\nu} F^{\mu\nu}$ that
appears in the Lagrangian density still scales as $\bar\rho_0^{-2/(D-2)}$.
Hence for large $\bar\rho_0$, locally the rolling black hole experiences almost flat
space-time.
 It also follows from this that a stationary observer in this background feels acceleration of order 
 $\bar\rho_0^{-1/(D-2)}$. 
In the sense described in footnote \ref{fo1}, this also gives 
an order of magnitude estimate
of the
temperature experienced by the rolling black hole / string.
From \refb{erhorange} we see
that this is much smaller than the temperature $T_{bh}\sim N^{-1/\{2(D-2)\}}$ 
given in \refb{e2.14}. Since \refb{erhorange} ensures that the effect of evaporation remains
small during
the transit, it now follows that the effect of absorption of the ambient
radiation also remains small during the transit.

We can also check that the effect of Hawking radiation from the background black
hole remains small. From \refb{ecanmet} we can estimate the Hawking temperature 
$T_{BH}$ 
of the background black hole to be of order $C^{-1} m^{(D-4)/(D-3)}$. After taking into account
the effect of the red-shift factor $(C\rho^{3-D})^{(D-3)/(D-2)}$ from \refb{ebackgroundlimit}, 
we get the red-shifted Hawking temperature at
$\rho\sim \bar\rho_0/C$ to be
\be
T_r \sim T_{BH}\times (C\rho^{3-D})^{(D-3)/(D-2)} \sim C^{D-4} \, m^{(D-4)/(D-3)}
\, \bar\rho_0^{-(D-3)^2 / (D-2)} << N^{-1/\{2(D-2)\}}\sim T_{bh}\, ,
\ee
where in the last step we have used \refb{ecdrel},
\refb{embound} and \refb{e2.40}. This shows that $T_r$ is small
compared to the temperature $T_{bh}$ of the rolling black hole. 
Since the effect of the latter is small, we conclude
that the effect of the absorption of 
Hawking radiation from the background black hole also remains small.


\subsection{Fall from the asymptotic region} \label{s2.4}

In sections \ref{s2.2} and \ref{s2.3} we have discussed how to make the black hole / string
roll over a
range where the effective string coupling $e^\phi$ varies from $N^{-1/4}/\eps$ 
to $N^{-1/4}\eps$.
In this section we shall explore 
whether the rolling black hole can roll over a range of $e^\phi$ that
starts at some small but $N$ independent value $\eta$ and then rolls all the way
to the region $e^\phi\sim N^{-1/4}\eps$ in a time that is small compared to the decay time of the
black hole. For this we have to start at a value $\bar\rho_1$ of $\bar\rho$ where $e^\phi\sim \eta$.
\refb{ebackgroundNEW} gives
\be \label{e225b}
C^{D-2}\bar\rho_1^{3-D}\sim \eta^{-2}\, .
\ee
Comparing this with \refb{ecdrel} we get,
\be\label{e237}
\bar\rho_1 \sim \bar\rho_0 \, (N\eta^4)^{1/\{2(D-3)\}}\, .
\ee
Note that $\bar\rho_0$ is still defined to be the value of $\bar\rho$ where $e^\phi\sim N^{-1/4}$.
Using \refb{ebackgroundNEW} we see that the 
total transit time in the frame of the rolling black hole 
from $\bar\rho\sim \bar\rho_1$ to $\bar\rho_0$  is of order
\be \label{e226b}
\bar\rho_1^{1/(D-2)}\sim \bar\rho_0^{1/(D-2)}\, (N\eta^4)^{1/\{2(D-3)(D-2)\}}\, ,
\ee
with most of the time spent in the region $\bar\rho\sim\bar\rho_1$.
Requiring this to be small compared to the evaporation time $\tau_{bh}\sim N^{(D-1)/\{2(D-2)\}}$
gives
\be \label{eupperN}
 \bar\rho_0 << \eta^{-2/(D-3)} N^{(D-1)/2 - 1/\{2(D-3)\}}\, .
\ee

The lower bound on $\bar\rho_0$ that arises from requiring that the rolling is adiabatic also gets
modified. For this we note from \refb{ebackgroundNEW} that 
 between $\bar\rho_1$ and $\bar\rho_0$ there is a redshift factor
\be \label{e238b}
\gamma\sim (\bar\rho_1/\bar\rho_0)^{(D-3)^2 / (D-2) }\sim (N\eta^4)^{(D-3)/ \{2 (D-2)\}}\, ,
\ee
where we used \refb{e237}. Hence a freely falling rolling black hole will have most of its
mass converted to kinetic energy by the time it falls to the position $\bar\rho_0$ and it will
move with ultra-relativistic speed.
Hence at $\bar\rho\sim\bar\rho_0$, 
the rolling black hole has time dilation by a factor of $\gamma$ and  the proper time of transit
from $e^\phi\sim N^{-1/4}/\eps$ to $e^\phi \sim N^{-1/4}\eps$ is given by $\bar\rho_0^{1/(D-2)}/\gamma$
instead of $\bar\rho_0^{1/(D-2)}$. For the adiabatic approximation to be valid, we need this to be
small compared to the size $r_s$ of the rolling black hole. This changes \refb{e2.36A} to 
\be 
\bar\rho_0^{1/(D-2)}(N\eta^4)^{-(D-3)/ \{2 (D-2)\}} >> N^{1 / \{2 (D-2)\}}\, .
\ee
This gives
\be \label{elowerN}
\bar\rho_0 >> N^{(D-2)/2}\,  \eta^{2(D-3)}\, .
\ee
There are various other
equivalent ways of arriving at the same conclusion, e.g. in the rest frame of the rolling
black hole, the
ambient geometry will have a length contraction by a factor of $\gamma$, 
and we need to demand that
even after taking this into account, the fields in the 
ambient space-time should vary over a length scale 
much larger than the size of the rolling black hole.

Combining \refb{elowerN} with \refb{eupperN} we get
\be
N^{(D-2)/2} \eta^{2(D-3)} << \bar\rho_0 << \eta^{-2/(D-3)} N^{(D-1)/2 - 1/\{2(D-3)\}}\, .
\ee
For $D\ge 5$ and large $N$, the upper bound is larger than the lower bound, and hence we can
find $\bar\rho_0$ satisfying these conditions. For $D=4$, the $N$ dependence of the two sides are
identical, but by taking $\eta$ to be small, we can still ensure that the upper bound is larger than the
lower bound. 
Thus we see
that it is possible for the rolling black hole to roll all the way from a region where the string coupling
is small but $N$ independent to the region where it becomes an elementary string excitation.
On the other hand,
if the rolling black hole undergoes a controlled fall as discussed in footnote \ref{fo1}, maintaining its
speed at order unity but not parametrically close to unity, then the lower bound
on $\bar\rho_0$ will be given by the earlier bound $N^{1/2}$, and even for $D=4$ and finite $\eta$, the upper
bound on $\bar\rho_0$ remains larger than the lower bound for large $N$.

Finally we can consider the fall from the asymptotic region $\rho\sim C^{1/(D-3)}$. In this case the transit time  is
of order $C^{1/(D-3)}$. Comparing this with the transit time $\bar\rho_1^{1/(D-2)}\sim C^{1/(D-3)}
\eta^{2/\{(D-2)(D-3)\}}$ computed from \refb{e226b} and \refb{e225b}, we see that this corresponds to setting 
$\eta\sim1$ in the previous analysis. Also  from \refb{ebackgroundlimit} and \refb{ecdrel} we see that 
the redshift factor $\gamma$ at a position $\rho\sim \bar\rho_0 / C$ 
is now of order 
$(C\rho^{3-D})^{(D-3)/(D-2)} \sim N^{(D-3)/\{2(D-2)\}}$.
Comparing this with \refb{e238b} we again see that that this corresponds to setting $\eta\sim 1$ in the
previous analysis. Thus for $D\ge 5$, even a free fall of the rolling black hole from the asymptotic region will
convert a black hole to an elementary string state in a controlled fashion. For $D=4$ the adiabatic approximation
will break down near the end of the roll, but this can be avoided by arresting the fall by external force as described
in footnote \ref{fo1} so that the  rolling black hole does not become ultra-relativistic during the roll.
In the extreme case, if we bring the rolling black hole at rest at $\bar\rho\sim \bar\rho_0\eps^{2/(D-3)}$
where it becomes an elementary string state, the proper acceleration felt  by the rolling black hole
will be of order $\bar\rho_0^{-1/(D-2)} \eps^{-2/\{(D-2)(D-3)\}}$. This remains small 
compared to the
string scale
as long as
\refb{e2.36A} holds. An observer in the rest frame of this black hole can study its quantum state and send
the information back to the asymptotic observer since both the rolling black hole and the observer
will be outside the global event horizon (which in the approximation we are using, is the horizon of
the background black hole at $\rho^{D-3}=2m$).

\sectiono{Other non-extremal black holes} \label{s3}

In this section we shall briefly discuss the case where the rolling black hole is 
a non-extremal black hole carrying charge(s) and / or angular momenta. We shall assume
that the black hole is finite distance away from extremality. In that case the scaling of
temperature, entropy and other thermodynamic quantities as a function of mass have the
same form as in the case of a Schwarzschild black hole, and the analysis follows a path that
is more or less identical to that in section \ref{s2}, except that the large parameter $N$ has to be
interpreted as the square of the entropy $S$ so that \refb{e1} holds:
\be
S\sim \sqrt N\, .
\ee
If $r_s$ denotes the horizon radius measured in the Planck scale, then we have
\be\label{e3.50}
r_s\sim S^{1/(D-2)}\sim N^{1/\{2(D-2)\}} \qquad \Rightarrow \qquad \wt r_s \sim g_s^{2/(D-2)}
N^{1/\{2(D-2)\}}\, ,
\ee
where we used \refb{etr2} in the second step. On the other hand,
it was shown in
\cite{9612146} that at the correspondence point, the size of the horizon, measured in string scale, 
is of
order unity, and hence \refb{erstring} holds:
\be\label{e3.51}
\wt r_s \sim 1\, .
\ee
Comparing \refb{e3.50} and \refb{e3.51},
we get
\be
g_s\sim  N^{-1/4}\, ,
\ee
at the correspondence point. This is the same relation as \refb{e2.9}. The rest of the analysis now
follows as in section \ref{s2}. The only other difference is that the rolling black hole is also a
source of the dilaton and other scalars and  the gauge fields,  
but for black holes that are not near extremal, 
the changes in various equations are of order unity and can be ignored.

For near extremal black holes the situation is more complicated since the temperature does not
scale as \refb{etbh} and we have a new large parameter. Since the effect of this is to increase the
evaporation time of the black hole, we expect that the upper bound on the transit time will be
larger and it will be easier to construct background in which the system transforms from a black hole
state to the regular quantum state. However, as pointed out in \cite{9612146}, 
for some near extremal systems
the correspondence principle has the puzzling feature that as we reduce the string coupling,
even before reaching the
correspondence point the black hole description breaks down since the curvature of the string
metric reaches the string scale in some region between the  horizon and the asymptotic region.
For this reason we shall not study these cases in detail. Instead in the next section we shall 
focus on the BPS black holes. 

\sectiono{BPS black holes} \label{s4}


BPS black holes have been used to count microstates of black holes to high precision. 
Since appropriate supersymmetric index of the black hole is expected to be protected under
a change in the coupling constant, we can carry out the counting at weak coupling where 
the system is described as an usual quantum state containing D-branes, fundamental
strings and other known objects, while at strong coupling the system is described as a
black hole\cite{9601029}. Hence by exploiting the varying dilaton background described in
section \ref{s2} we can make the transition between the black hole and microscopic description
dynamical. As the black hole rolls from the strong to the weak coupling region, its description
changes from that of a regular extremal BPS black hole to a system containing D-branes and other
objects with regular quantum mechanical description.

In fact, for BPS black holes the situation is better than that for non-BPS black holes discussed in
the previous sections since the BPS black holes have zero temperature and do not evaporate.
Hence there is no upper limit on the time of rolling. By taking the scaling parameter 
$\lambda$ of section \ref{s2.2} 
(or equivalently the parameter $C$ of the background black hole of section \ref{s2.3}) to be 
sufficiently large one can ensure that a not a single quantum of ambient Hawking
or Unruh radiation is
absorbed by the rolling black hole and hence the state of the rolling black hole remains
unchanged during the motion. This can be seen by noting that the ambient temperature of the
background scales as
$\lambda^{-1}$ and hence the ambient entropy density scales as $\lambda^{-(D-1)}$. This has to
be multiplied by the travel time $\lambda$ and the area of the rolling black hole to compute the
total entropy absorbed during the transit. Since the area of the rolling black hole does not scale
with $\lambda$, we see that the net entropy absorbed during the transit scales as
$\lambda^{-(D-2)}$ and can be made as small as we like by taking $\lambda$ large. In particular
when it becomes much smaller than one, it will imply that not even a single quantum of
ambient radiation is absorbed by the
black hole during its motion.

As a specific example, we can consider the Strominger-Vafa black holes\cite{9601029}.
They are black holes in type IIB string theory  compactified on $\MM\times S^1$, where $\MM$
can be either $K3$ or $S^1$, carrying charges corresponding to $Q_5$ D5-branes wrapped on
$\MM\times S^1$, $Q_1$ D1-branes wrapped on $S^1$ and $N$ units of momenta along $S^1$.
We shall take $Q_5\sim Q_1\sim N$ and denote by $g_5$ the asymptotic value of the type IIB
string coupling. For sufficiently small $g_5$, 
the system can be described as an ordinary quantum system made
of D-branes and momenta and the counting of states can be done in this regime. On the other hand
for sufficiently large $g_5$ 
the more appropriate description of the system is as an extremal black hole.
Hence we need a background in which $g_5$ varies from a relatively larger value to a small value. 
This can be
achieved by the same background black hole that was used in section \ref{s2.3}.

\sectiono{Discussion} \label{s6}

In our analysis we have shown that the black hole can be transformed into an ordinary quantum
system containing stings and branes
by letting it drop to a region close to a much bigger near
extremal electrically
charged black hole. We have not discussed how to extract information from this
system, but since this is an ordinary quantum system, an observer falling with the rolling black
hole should be able to study the quantum state of this system using standard methods and
transmit the data to an asymptotic observer. The time
scale of such experiments will be long due to the weakness of the string coupling and one might
wonder whether the rolling black hole will fall into the background black hole before such experiments
could be performed. 
Furthermore as the coupling becomes weaker as we approach the horizon of the background
black hole the time scale of measurement becomes longer. 
The latter problem could be avoided by adding a small amount of magnetic
charge to the background black hole so that the attractor value of the string coupling is given by
$N^{-1/4}/\eps$ for some fixed small number $\eps$. In that case,
even if the rolling black hole falls
towards the horizon of the background black hole, the coupling does not decrease any further.
Also one could in principle arrest the fall of the rolling black hole towards
the background black hole using the classical gravitational
field of other (accelerating) objects placed in appropriate regions of space-time (see footnote
\ref{fo1} for a discussion on this).. 
While we have
not studied this problem in detail, we expect that this should be possible in principle as long as
the system of which the rolling black hole is a part 
remains outside the horizon of the background black hole.

Since black holes are expected to have chaotic spectrum while the spectrum of weakly coupled
strings and
branes have regular pattern, one might wonder how the transition from black hole to the brane
description affects the spectrum. 
To this end, note that the spectrum of any microscopic system made of
branes and strings is renormalized by quantum corrections.
If the quantum corrections are chaotic, this could turn a regular spectrum into a chaotic spectrum.
In particular it was shown in \cite{2103.15301} that  the interaction between the
string states is highly chaotic. Since the renormalized mass depends on the interaction, it is quite
plausible that the renormalized masses will display a chaotic spectrum.

\bigskip

\noindent{\bf Acknowledgement:} 
I wish to thank 
Nejc Ceplak, Roberto Emparan, Andrea Puhm,  Marija Tomasevic and Yoav Zigdon for 
useful communications.
This work was supported by the ICTS-Infosys Madhava 
Chair Professorship
and the Department of Atomic Energy, Government of India, under project no. RTI4019.

\appendix

\sectiono{Electrically charged black hole solution} \label{sa}

In this appendix we shall review the construction of the electrically charged,
non-rotating black hole solution of string theory in $D$ non-compact space-time dimensions.
First we shall consider heterotic string theory compactified on a six dimensional
torus $T^6$, and then show how the same construction can be generalized to any string
compactification for which the compact space has a circle factor.
The massless fields of this theory consist of the canonical
metric $ds^2$, the two form field, the four dimensional dilaton $\phi$, a set of
scalars taking values in $O(6,22)/O(6)\times O(22)$,
encoded in a symmetric $O(6,22)$ matrix $M$ and 28 gauge fields, collectively denoted
by a 28 dimensional vector $A_\mu$. We denote by $\wt{ds}^2=e^{2\phi} ds^2$ the string metric.
The background black hole will be taken to be an 
electrically charged non-rotating black hole solution, given by\cite{9411187}
\ben \label{e2.1app}
\wt{ds}^2 &=& -\Delta^{-1} \, \rho^2 (\rho^2-2m\rho) dt^2 + \rho^2 \, (\rho^2-2m\rho)^{-1} d\rho^2
+\rho^2 \,  d\Omega_2^2 \, , \nonumber \\
e^{-2\phi} &=&  \Delta^{1/2} \, \rho^{-2} \, , \nonumber \\
M &=& I_{28} + \pmatrix{Pnn^T & Q n p^T \cr Q p n^T & Ppp^T} \, , \nonumber \\
A_t &=& -{1\over \sqrt 2} \, m\, \rho\, \Delta^{-1} \, \pmatrix{
\sinh\alpha \{\cosh\beta \rho^2
+m\rho(\cosh\alpha-\cosh\beta)\} \vec n \cr
\sinh\beta \{\cosh\alpha \rho^2
+m\rho(\cosh\beta-\cosh\alpha)\} \vec p
}\, ,
\een
where
\ben \label{edeldef}
\Delta &\equiv& \rho^4 + 2 m\rho^3(\cosh\alpha\cosh\beta-1) + m^2 \rho^2 
(\cosh\alpha-\cosh\beta)^2  \, , \nonumber \\
P &\equiv & 2\, \Delta^{-1} \, m^2 \rho^2 \sinh^2\alpha\sinh^2\beta\, , \nonumber \\
Q &\equiv & - 2\Delta^{-1}m\rho  \sinh\alpha\sinh\beta \{ \rho^2 + m\rho (\cosh\alpha\cosh\beta
-1)\}
\, ,
\een
$\vec n$ is a 22 dimensional unit vector and $\vec p$ is a six dimensional unit vector.
The 2-form field vanishes. $m$, $\alpha$ and $\beta$ are arbitrary constants, that, together
with $\vec n$ and $\vec p$, label the mass and the charges carried by this 
background black hole.

We set 
\be
\alpha=0, \qquad m(\cosh\beta-1) = C\, ,
\ee
so that we have
\be
\Delta = \rho^4 \left( 1 + C\rho^{-1}\right)^2, \qquad P=0, \qquad Q=0\, .
\ee
Then \refb{e2.1app}
takes the form:
\ben \label{ebackground}
\wt{ds}^2 &=& - (1+ C\rho^{-1})^{-2} (1-2m\rho^{-1}) \, dt^2 + (1-2m\rho^{-1})^{-1}
 d\rho^2
+ \rho^2 d\Omega_2^2 \, , \nonumber \\
e^{-2\phi} &=&  (1+ C\rho^{-1}) 
\, , \nonumber \\
M &=& I_{28}\, , \nonumber \\
A_t &=& -{1\over \sqrt 2} \, m\, \sinh\beta\, \rho^{-1}\, (1+ C\rho^{-1})^{-1}\, \pmatrix{
\vec 0 \cr
\vec p
}\, .
\een

Physically $M$ denotes the moduli associated with the
components of the metric, the two form
field and the ten dimensional gauge fields along $T^6$ and $A_\mu$ denotes the 
four dimensional gauge fields associated with the components of the metric and two form field
with one index along internal direction and one index along the non-compact direction and
the components of the ten dimensional gauge field along the non-compact directions. In particular
the solution described above carries equal amount of 
fundamental string winding and momentum charges  along the direction
$\vec p$ in $T^6$.\footnote{The equality of momentum and winding charges is not necessary for
our analysis. For any other finite ratio of these charges, $M$ would approach some other constant
matrix at small $\rho$.}

From this description it is clear that the same solution can be lifted to 
any compactification to 3+1 dimensions
in which we have one circle direction among the compact coordinates. $M$ is in any case
frozen to a constant and we simply have to interpret $A_t$ as the gauge field that couples to
both momentum and winding charge along the circle. To lift it to a compactification with $D$
non-compact space-time dimensions with at least 
one circle among the compact coordinates,  we need
to recall how the solution was constructed in \cite{9411187}. The essential idea was to
start with a Schwarzschild black hole in four space-time dimensions and then perform a
duality rotation that mixed the time direction with the compact 
directions\cite{oddone,9108011,9109038}. The same
operation can be carried out in $D$ space-time dimensions by starting with a Schwarzschild
black hole in $D$ space-time dimensions. The only difference will be that in the initial
solution all factors of $m/\rho$ would be replaced by $m/\rho^{D-3}$ and $d\Omega_2^2$
will be replaced by $d\Omega_{D-2}^2$. This will give the solution for $D$ 
non-compact space-time
directions\cite{9506200}:
\ben \label{ebackgroundD}
\wt{ds}^2 &=& - (1+ C\rho^{3-D})^{-2} (1-2m\rho^{3-D})\,  dt^2 + (1-2m\rho^{3-D})^{-1}
 d\rho^2
+ \rho^2 d\Omega_{D-2}^2 \, , \nonumber \\
e^{-2\phi} &=&  (1+ C\rho^{3-D}) 
\, , \nonumber \\
A_t &=& -{1\over \sqrt 2} \, m\, \sinh\beta\, \rho^{3-D}\, (1+ C\rho^{3-D})^{-1} \, .
\een
We have dropped the expression for $M$ since this is a constant anyway, and also
not displayed the internal vector $\vec p$
associated with the gauge field, with the understanding that
we are considering a solution carrying equal amount of momentum and winding charges
along one of the compact circles. The canonical $D$ dimensional metric $ds^2=
e^{-4\phi/(D-2)} \wt{ds}^2$ now takes the form:
\ben
ds^2 &=& - (1+ C\rho^{3-D})^{-2(D-3)/(D-2)} (1-2m\rho^{3-D}) dt^2 + 
(1+ C\rho^{3-D})^{2/(D-2)} (1-2m\rho^{3-D})^{-1}
 d\rho^2\nonumber \\
&+& (1+ C\rho^{3-D})^{2/(D-2)} \, \rho^2\, d\Omega_{D-2}^2 \, .
\een


\begin{thebibliography}{99}

\bibitem{9612146}
G.~T.~Horowitz and J.~Polchinski,
``A Correspondence principle for black holes and strings,''
Phys. Rev. D \textbf{55} (1997), 6189-6197
doi:10.1103/PhysRevD.55.6189
[arXiv:hep-th/9612146 [hep-th]].

\bibitem{9707170}
G.~T.~Horowitz and J.~Polchinski,
``Selfgravitating fundamental strings,''
Phys. Rev. D \textbf{57} (1998), 2557-2563
doi:10.1103/PhysRevD.57.2557
[arXiv:hep-th/9707170 [hep-th]].

\bibitem{9907030}
T.~Damour and G.~Veneziano,
``Selfgravitating fundamental strings and black holes,''
Nucl. Phys. B \textbf{568} (2000), 93-119
doi:10.1016/S0550-3213(99)00596-9
[arXiv:hep-th/9907030 [hep-th]].

\bibitem{bowick}
M.~J.~Bowick, L.~Smolin and L.~C.~R.~Wijewardhana,
``Role of String Excitations in the Last Stages of Black Hole Evaporation,''
Phys. Rev. Lett. \textbf{56} (1986), 424
doi:10.1103/PhysRevLett.56.424

\bibitem{9309145}
L.~Susskind,
``Some speculations about black hole entropy in string theory,''
[arXiv:hep-th/9309145 [hep-th]].

\bibitem{9605112}
E.~Halyo, A.~Rajaraman and L.~Susskind,
``Braneless black holes,''
Phys. Lett. B \textbf{392} (1997), 319-322
doi:10.1016/S0370-2693(96)01544-4
[arXiv:hep-th/9605112 [hep-th]].

\bibitem{9609075}
E.~Halyo, B.~Kol, A.~Rajaraman and L.~Susskind,
``Counting Schwarzschild and charged black holes,''
Phys. Lett. B \textbf{401} (1997), 15-20
doi:10.1016/S0370-2693(97)00357-2
[arXiv:hep-th/9609075 [hep-th]].

\bibitem{2109.08563}
Y.~Chen, J.~Maldacena and E.~Witten,
``On the black hole/string transition,''
JHEP \textbf{01} (2023), 103
doi:10.1007/JHEP01(2023)103
[arXiv:2109.08563 [hep-th]].

\bibitem{2107.09001}
R.~Brustein and Y.~Zigdon,
``Black hole entropy sourced by string winding condensate,''
JHEP \textbf{10} (2021), 219
doi:10.1007/JHEP10(2021)219
[arXiv:2107.09001 [hep-th]].

\bibitem{2205.15976}
Y.~Matsuo,
``Fluid model of a black hole-string transition,''
Phys. Rev. D \textbf{107} (2023) no.12, 126003
doi:10.1103/PhysRevD.107.126003
[arXiv:2205.15976 [hep-th]].



\bibitem{2307.03573}
N.~{\v{C}}eplak, R.~Emparan, A.~Puhm and M.~Toma{\v{s}}evi{\'c},
``The correspondence between rotating black holes and fundamental strings,''
JHEP \textbf{11} (2023), 226
doi:10.1007/JHEP11(2023)226
[arXiv:2307.03573 [hep-th]].



\bibitem{2502.07883}
A.~Sen,
``Are Moduli Vacuum Expectation Values or Parameters?,''
[arXiv:2502.07883 [hep-th]].

\bibitem{2503.00601}
A.~Sen,
``How to Create a Flat Ten or Eleven Dimensional Space-time in the Interior of an Asymptotically Flat Four Dimensional String Theory,''
[arXiv:2503.00601 [hep-th]].

\bibitem{2506.13876}
A.~Sen,
``Decorating Asymptotically Flat Space-Time with the Moduli Space of String Theory,''
[arXiv:2506.13876 [hep-th]].

\bibitem{2501.17697}
T.~Banks,
``Old Ideas for New Physicists III: String Theory Parameters are NOT Vacuum Expectation Values,''
[arXiv:2501.17697 [hep-th]].


\bibitem{Hawking:1976ra}
S.~W.~Hawking,
``Breakdown of Predictability in Gravitational Collapse,''
Phys. Rev. D \textbf{14} (1976), 2460-2473
doi:10.1103/PhysRevD.14.2460

\bibitem{Page:1993wv}
D.~N.~Page,
``Information in black hole radiation,''
Phys. Rev. Lett. \textbf{71} (1993), 3743-3746
doi:10.1103/PhysRevLett.71.3743
[arXiv:hep-th/9306083 [hep-th]].

\bibitem{0502050}
S.~D.~Mathur,
``The Fuzzball proposal for black holes: An Elementary review,''
Fortsch. Phys. \textbf{53} (2005), 793-827
doi:10.1002/prop.200410203
[arXiv:hep-th/0502050 [hep-th]].

\bibitem{0909.1038}
S.~D.~Mathur,
``The Information paradox: A Pedagogical introduction,''
Class. Quant. Grav. \textbf{26} (2009), 224001
doi:10.1088/0264-9381/26/22/224001
[arXiv:0909.1038 [hep-th]].

\bibitem{Penington:2019npb}
G.~Penington,
``Entanglement Wedge Reconstruction and the Information Paradox,''
JHEP \textbf{09} (2020), 002
doi:10.1007/JHEP09(2020)002
[arXiv:1905.08255 [hep-th]].

\bibitem{Almheiri:2019psf}
A.~Almheiri, N.~Engelhardt, D.~Marolf and H.~Maxfield,
``The entropy of bulk quantum fields and the entanglement wedge of an evaporating black hole,''
JHEP \textbf{12} (2019), 063
doi:10.1007/JHEP12(2019)063
[arXiv:1905.08762 [hep-th]].

\bibitem{1908.10996}
A.~Almheiri, R.~Mahajan, J.~Maldacena and Y.~Zhao,
``The Page curve of Hawking radiation from semiclassical geometry,''
JHEP \textbf{03} (2020), 149
doi:10.1007/JHEP03(2020)149
[arXiv:1908.10996 [hep-th]].

\bibitem{Almheiri:2020cfm}
A.~Almheiri, T.~Hartman, J.~Maldacena, E.~Shaghoulian and A.~Tajdini,
``The entropy of Hawking radiation,''
Rev. Mod. Phys. \textbf{93} (2021) no.3, 035002
doi:10.1103/RevModPhys.93.035002
[arXiv:2006.06872 [hep-th]].

\bibitem{Laddha:2020kvp}
A.~Laddha, S.~G.~Prabhu, S.~Raju and P.~Shrivastava,
``The Holographic Nature of Null Infinity,''
SciPost Phys. \textbf{10} (2021) no.2, 041
doi:10.21468/SciPostPhys.10.2.041
[arXiv:2002.02448 [hep-th]].




\bibitem{9707207}
E.~Cremmer, H.~Lu, C.~N.~Pope and K.~S.~Stelle,
``Spectrum generating symmetries for BPS solitons,''
Nucl. Phys. B \textbf{520}, 132-156 (1998)
doi:10.1016/S0550-3213(98)00057-1
[arXiv:hep-th/9707207 [hep-th]].

\bibitem{Unruh:1976db}
W.~G.~Unruh,
``Notes on black hole evaporation,''
Phys. Rev. D \textbf{14} (1976), 870
doi:10.1103/PhysRevD.14.870

\bibitem{9601029}
A.~Strominger and C.~Vafa,
``Microscopic origin of the Bekenstein-Hawking entropy,''
Phys. Lett. B \textbf{379} (1996), 99-104
doi:10.1016/0370-2693(96)00345-0
[arXiv:hep-th/9601029 [hep-th]].


\bibitem{2103.15301}
D.~J.~Gross and V.~Rosenhaus,
``Chaotic scattering of highly excited strings,''
JHEP \textbf{05} (2021), 048
doi:10.1007/JHEP05(2021)048
[arXiv:2103.15301 [hep-th]].



\bibitem{9411187}
A.~Sen,
``Black hole solutions in heterotic string theory on a torus,''
Nucl. Phys. B \textbf{440} (1995), 421-440
doi:10.1016/0550-3213(95)00063-X
[arXiv:hep-th/9411187 [hep-th]].

\bibitem{oddone}
A.~Sen,
``O(d) x O(d) symmetry of the space of cosmological solutions in string theory, scale factor duality and two-dimensional black holes,''
Phys. Lett. B \textbf{271} (1991), 295-300
doi:10.1016/0370-2693(91)90090-D

\bibitem{9108011}
A.~Sen,
``Twisted black p-brane solutions in string theory,''
Phys. Lett. B \textbf{274} (1992), 34-40
doi:10.1016/0370-2693(92)90300-S
[arXiv:hep-th/9108011 [hep-th]].

\bibitem{9109038}
S.~F.~Hassan and A.~Sen,
``Twisting classical solutions in heterotic string theory,''
Nucl. Phys. B \textbf{375} (1992), 103-118
doi:10.1016/0550-3213(92)90336-A
[arXiv:hep-th/9109038 [hep-th]].

\bibitem{9506200}
A.~W.~Peet,
``Entropy and supersymmetry of D-dimensional extremal electric black holes versus string states,''
Nucl. Phys. B \textbf{456} (1995), 732-752
doi:10.1016/0550-3213(95)00537-2
[arXiv:hep-th/9506200 [hep-th]].


\end{thebibliography}
\end{document}